\documentclass[useAMS,usenatbib]{mn2e}

\usepackage[dvips]{graphicx}
\usepackage[T1]{fontenc}
\usepackage{ae,aecompl}

\title[Antenna Performance Analysis]{Antenna Performance Analysis for
Decameter Solar Radio Observations}
\author[A.A. Stanislavsky et al.]
       {A.A. Stanislavsky\thanks{E-mail: alexstan@ri.kharkov.ua}$^{1}$, A.A. Konovalenko$^{1}$, E.P. Abranin$^{1}$, V.V. Dorovskyy$^{1}$,
       \newauthor V.N. Mel'nik$^{1}$, M.L. Kaiser$^{2}$, A. Lecacheux$^{3}$, H. O. Rucker$^{4}$\\
        $^1$ Institute of Radio Astronomy, 4 Chervonopraporna St., Kharkov, 61002
        Ukraine\\
        $^2$ NASA GSFC, Code 674, Greenbelt, MD 20771, USA\\
        $^3$ Departement de Radioastronomie CNRS UMR 8644, Observatoire de Paris,
        France\\
        $^4$  Space Research Institute, Austrian Academy of Sciences, Schmiedlstrasse 6, A-8042, Graz, Austria}

\date{Received}

\pagerange{\pageref{firstpage}--\pageref{lastpage}} \pubyear{2009}

\begin{document}

\maketitle

\label{firstpage}

\begin{abstract}
Decameter wavelength radio emission is finely structured in solar bursts. For their
research it is very important to use a sufficient sensitivity of antenna systems. In this
paper we study an influence of the radiotelescope-antenna effective area on the results
of decameter solar radio observations. For this purpose we compared the solar bursts
received by the array of 720 ground-based dipoles and the single dipole of the
radiotelescope UTR-2. It's shown that a larger effective area of the ground-based antenna
allows us to measure a weaker solar emission and to distinguish a fine structure of
strong solar events. This feature has been also verified by simultaneous ground- and
space-based observations in the overlapping frequency range.
\end{abstract}


\section{Introduction}
The astronomy and radio astronomy investigations by means of spacecrafts and space
missions have extraordinary importance for science and mankind, and nobody calls in
question now. It is enough to recall about the famous space telescope of Hubble that gave
and gives a huge amount of new scientific results. Really, its location behind the
terrestrial atmosphere allowed one to take off restrictions due to optic effects of the
atmosphere and to open a new page in the exploration of the Universe. In this connection
a natural wish to reach the same success by spacecrafts in other frequency ranges has
arisen. One of challenges, sufficiently difficult for exploring from the Earth, is the
decameter wavelength emission. Although radioastronomy was born at low frequency of 20.5
MHz in the 1930s, it rapidly was developed towards higher frequencies for higher
resolution and better sensitivity. Great many disturbances and interference from various
radio broadcast stations, propagated world-wide in this frequency range, plus undesirable
and not-well-predictable ionospheric effects pulled up the progress of decameter radio
astronomy, though it would answer many very interesting astrophysical problems. However,
to launch the same (in the sense of efficiency as the telescope of Hubble and similar
others) decameter radiotelescope in the near-earth space (or for example, even in the
backside of the Moon, just some overbold heads proposed; see, for example, Smith 1990;
Takahashi 2003) is not quite simple. The point is that the receiving antenna capability
of such a radio telescope depends, in many respects, on its effective area that, for its
turn, depends on a wavelength of the receiving emission. If it is large (in particular,
decameters in our case), then the antenna sizes should be corresponding. It is clear that
the facilities of space technology are not boundless. Therefore, in the present day the
space missions (WIND, STEREO and others), carried out radio observations in the decameter
wavelength range, can permit themselves only some dipoles on their board, as the size of
such spacecrafts and their weight are restricted. How effective is the telescope it? Of
course, it is not comparable with the capability of the space telescope as Hubble, etc.
But it would be useful to estimate its facilities in comparison with ground-based
instruments such as, for example, the well-known biggest decameter radiotelescope UTR-2.
The aim of this paper is to answer this posed question.

For this purpose we have carried out simultaneously the radio astronomy observations in
the radiotelescope UTR-2 by two different antennas in the same frequency band. One of
them was an array of 720 dipoles, and another consisted of one. With all this going on
the concrete problem was on the agenda: to examine what intensity of solar bursts one
would be able to detect in the first and the second case. The other important question,
close to the previous one, is that how much (and what?) it is possible to distinguish a
fine structure of solar radio bursts in both cases. Next, we have compared the ordinary
simultaneous observations, carried out by the array of 720 dipoles (a part of the
radiotelescope UTR-2) and the Waves WIND instruments. This allows us to establish a real
effectiveness of space radiotelescopes in the study of weak solar events.

\section{Ground-based observations by two antennas}
Usually the observations of solar radio emission in the framework of the telescope UTR-2
are carried out by four banks of the antenna arm ``North'' or ``South'' (see the details
about the instrument in Braude et~al. 1978). In this case the antenna directional pattern
covers all the size of the solar upper corona from which the decameter radio bursts are
generated. These four banks are grouped into an antenna array of 720 dipoles
($A_{e}$~=~900 $\times$ 52 m$^2$), permitting one to receive a signal in the continuous
frequency band from 9 to 31 MHz. As a recording back-end, we use a two-channel discrete
spectrum analyzer (DSP). Its features are described in the paper of Mel'nik et~al.
(2004). We recorded simultaneously solar signals from the two antennas. The first antenna
had four banks of the antenna arm ``South'', and the second contained only one dipole of
the antenna arm ``North''. The observations were obtained between 15 and 27 August 2007
during three hours forenoon and afternoon. The solar activity was weak enough.
Nevertheless, we have observed some weak solar events. The tracking of the solar motion
in the sky by the antenna directional pattern of UTR-2 is performed electronically by
switching the phase-shift network of the telescope antenna. Antenna steering and data
acquisition are computer controlled. The observation range was between 18.5 and 30.5 MHz.
The calibration was carried out by the noise generator with the well-known spectral
density for both antenna channels simultaneously.

Since the back end in the solar observations was the same for both antennas, it is
convenient to choose a measure of their sensitivity in the following form:
\begin{equation}
p=T_{\rm A}/S_o=A_{\rm e}\times 10^{-22}/2k\quad {\rm Kelvins/sfu},\label{eq1}
\end{equation}
where $T_{\rm A}$ is the equivalent antenna temperature, $S_o$ the flux density in sfu (1
sfu = 10$^{-22}$ W m$^{-2}$ Hz$^{-1}$), $k$ the Boltzmann's constant. Four banks of the
antenna arm ``South'' gives a sensitivity of $\approx$~170,000 K/sfu, and a single dipole
(receiving over a perfect conducting ground plane) has only $\approx$ 228 K/sfu at 18.5
MHz, falling to $\approx$ 84 K/sfu at 30.5 MHz.

\begin{figure*}
\includegraphics[width=17cm]{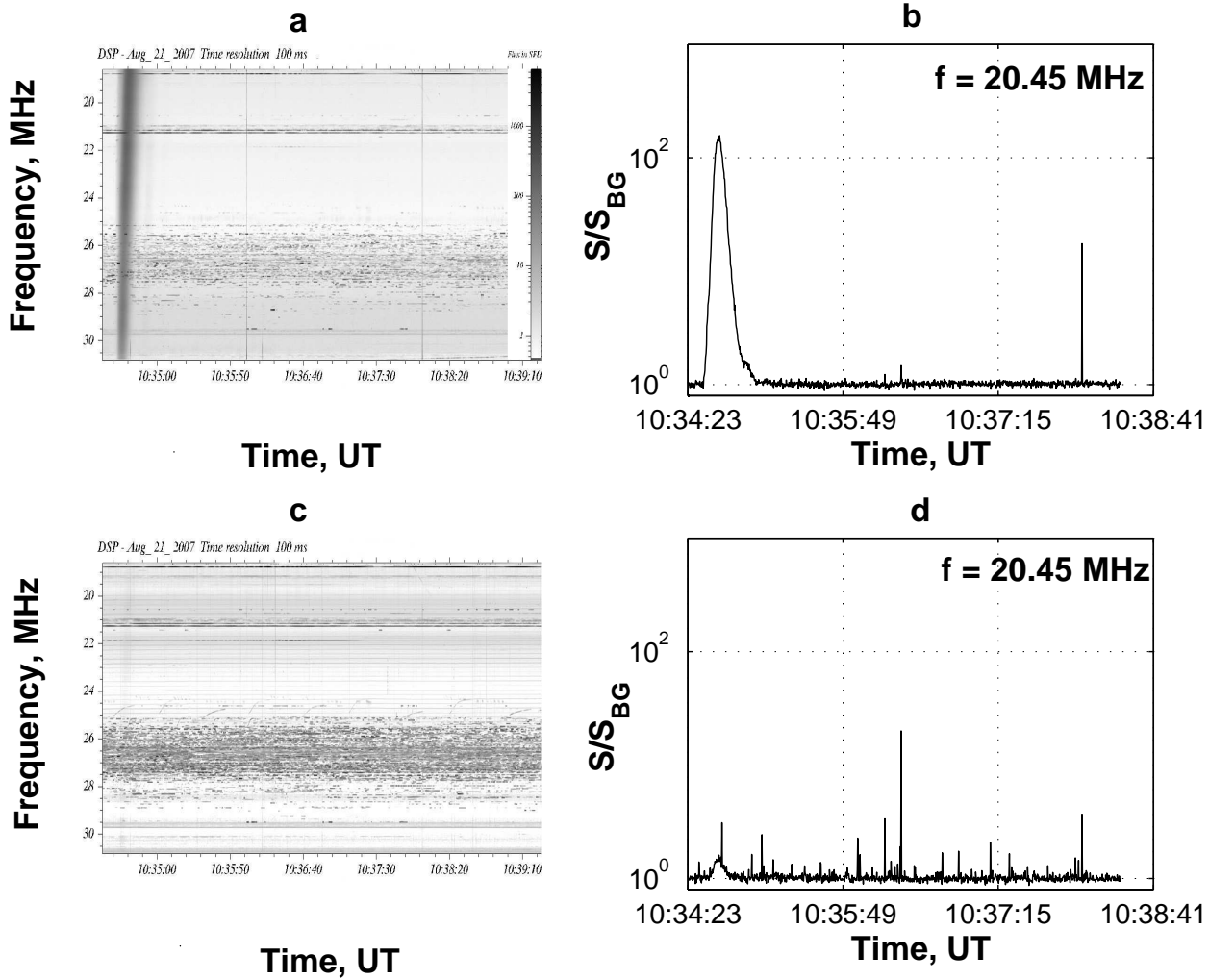}
 \caption{Single solar III-type burst has about 100 sfu in intensity (2007 August 21): top panels
correspond to measurements by the antenna array of 720 dipoles; bottom panels are results
of receiving the emission by one dipole (both records are simultaneous). The right panels
show the same profile of the burst, measured by the antennas in the relative units to the
background-flux-density level.\label{fig1}}
\end{figure*}

One of these records is represented in Fig. \ref{fig1}. The top panel shows the
spectrogram, received by the array of 720 dipoles, and the bottom panel demonstrates a
dynamical spectrum of emission received by only one dipole. About 10:34:35 UT in 2007
August 21 we have detected a single type III solar burst with a weak flux. From these
pictures it is clearly seen that the detection of such solar bursts is hardly possible
for one dipole. The Wind spectrum records do not show any burst at this time too (see
ftp://stereowaves.gsfc.nasa.gov/wind\_rad2/rad2pdf/). Probably, the burst flux density
was less than the minimum detectable value of the instrument.

\begin{figure*}
\includegraphics[width=17cm]{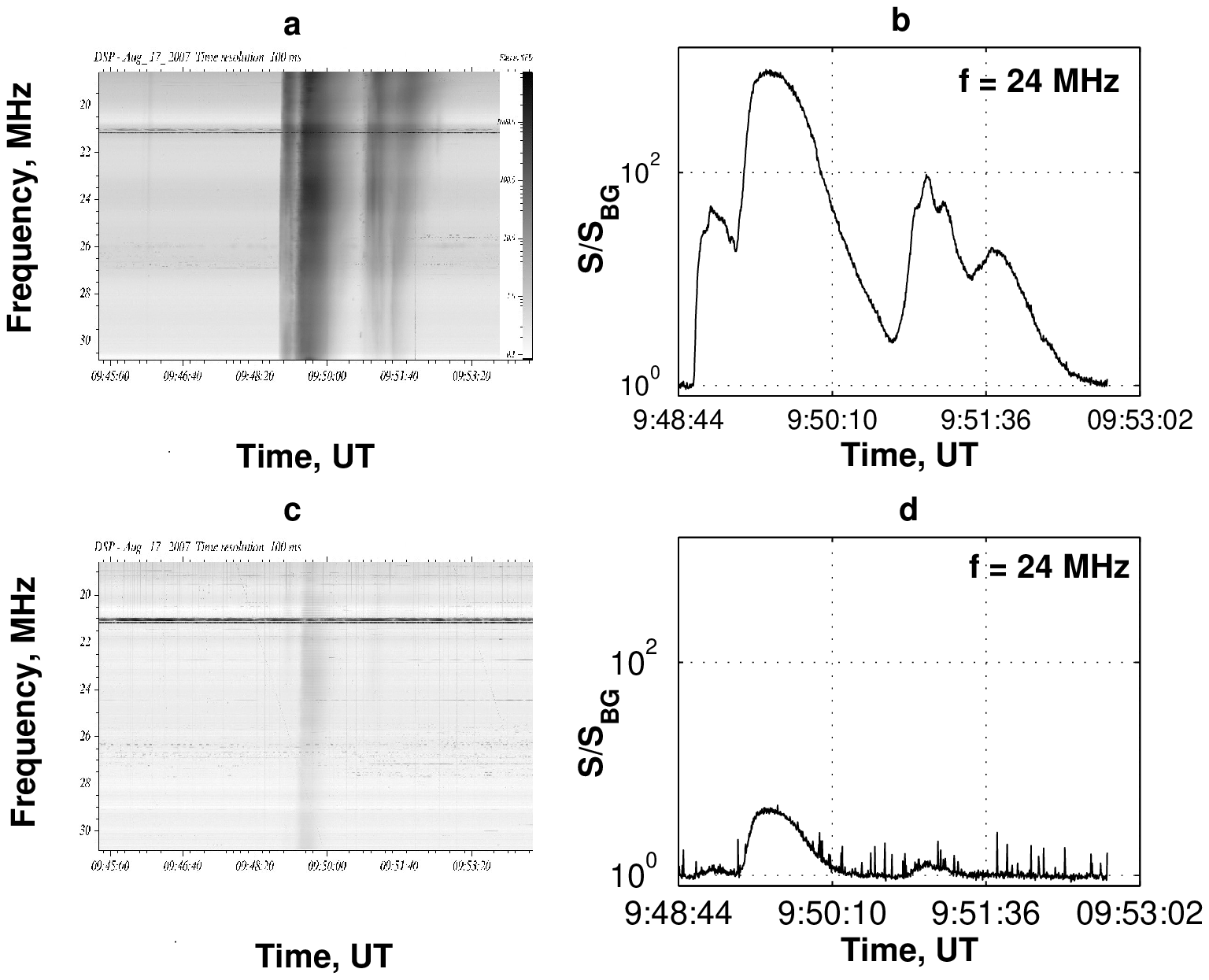}
 \caption{Set of bursts (2007 August 17) with various features in intensity: top left panel is a
spectrogram, obtained by the antenna array of 720 dipoles; bottom left panel demonstrates
a result of receiving by one dipole (both observations are simultaneous). On the right
the top and bottom panels are profiles of this event in the relative units to the
background-flux-density level.\label{fig2}}
\end{figure*}

The other restriction, leaving traces on the observations by one dipole, is that its
performance does not permit one to study a fine structure of the decameter bursts in
detail. As a significant example, let us consider Fig. \ref{fig2}. It presents a
spectrogram of a set of solar bursts with different features in intensity, that were
observed in 2007 August 17 at about 9:50 UT. The spectrogram, obtained by one dipole,
shows only one strongest burst, and it just plain didn't look like the set of bursts that
represented in the top panels. In contrast to the burst, represented in Fig. \ref{fig1},
for which the intensity and the duration increased uniformly with the decay of frequency,
the peculiarity of the given set of bursts is that their intensity and duration were
irregular in frequency. However, this can be tracked more or less successfully in the top
panels of Fig. \ref{fig2}, which give a sufficient resolution of this event in intensity
owing to the considerable dominance of the array in effective area. The Wind spectrum
also cannot resolve any fine structure of this event, although the strongest burst is
noticeable. To resume as a preliminary, we should tell that the antenna array of 720
dipoles has a considerable advantage for the study of weak solar bursts as well as for
the analysis of a fine structure of strong ones. One more interesting problem is that, as
is known, the solar type III bursts observed at metric wavelengths often disappear in
decametric range, whereas other bursts appear at decametric wavelengths and may or may
not continue to the low-frequency limit of most ground-based spectrographs at frequencies
less than 20-30 MHz. Applying the large decameter radioastronomy for this study, it would
be useful also.

\section{Space observations by the Waves WIND instrument at 9-13.8 MHz}
The spacecraft WIND is located in the Lagrangian point L1 between the Earth and the Sun,
where the gravitational attraction of this spacecraft to the Earth becomes balanced with
the gravitation attraction of the spacecraft to the Sun. This point is situated about 1.5
million km from the Earth. However, the distance from the spacecraft to the Sun is more
in 99 times. To recall one of formulae from Kraus (1967), the flux density of radio
emission observed at the distance $r$ is inversely as the square of the distance.
Therefore, the antenna of the spacecraft WIND receives a signal only in 2$\%$ greater
than the same antenna from the Earth, if we do not account for scattering, absorption and
rejection of radio emission in the terrestrial ionosphere and the near-Earth space.
However, the increase of the antenna effective area for any spacecraft is problematic
enough because of the causes discussed above in Introduction, whereas for any
ground-based antenna this is quite real (by building an antenna array). What does this
give? The spectral power, received by antenna, is directly proportional to its effective
area. This means that the array of 720 dipoles increase a spectral density of received
signals proportionally so many times as its effective area is more in comparison with one
dipole. In addition to that it should be added an appreciable advance of interference
protection in many times. This problem quite relates to the one-dipole antenna of the
spacecraft WIND (or STEREO and others), if the antenna is compared with ground-based
instruments such as the radiotelescope UTR-2 in the form of the array of 720 dipoles
applied for solar observations. Both instruments have a piece of the frequency range,
overlapping about 9-13.8 MHz. Here we can compare their capability in the solar burst
study from the real radio astronomy observations. In this case the delay between
simultaneous records of WIND and UTR-2 because of different spatial locations is only
about 5 s, but to detect the feature from the records themselves is impossible, as 256
frequency channels are sequentially switched in the receiver RAD2 of the spacecraft WIND
during 16.192 s.

\begin{figure*}
\includegraphics[width=15cm]{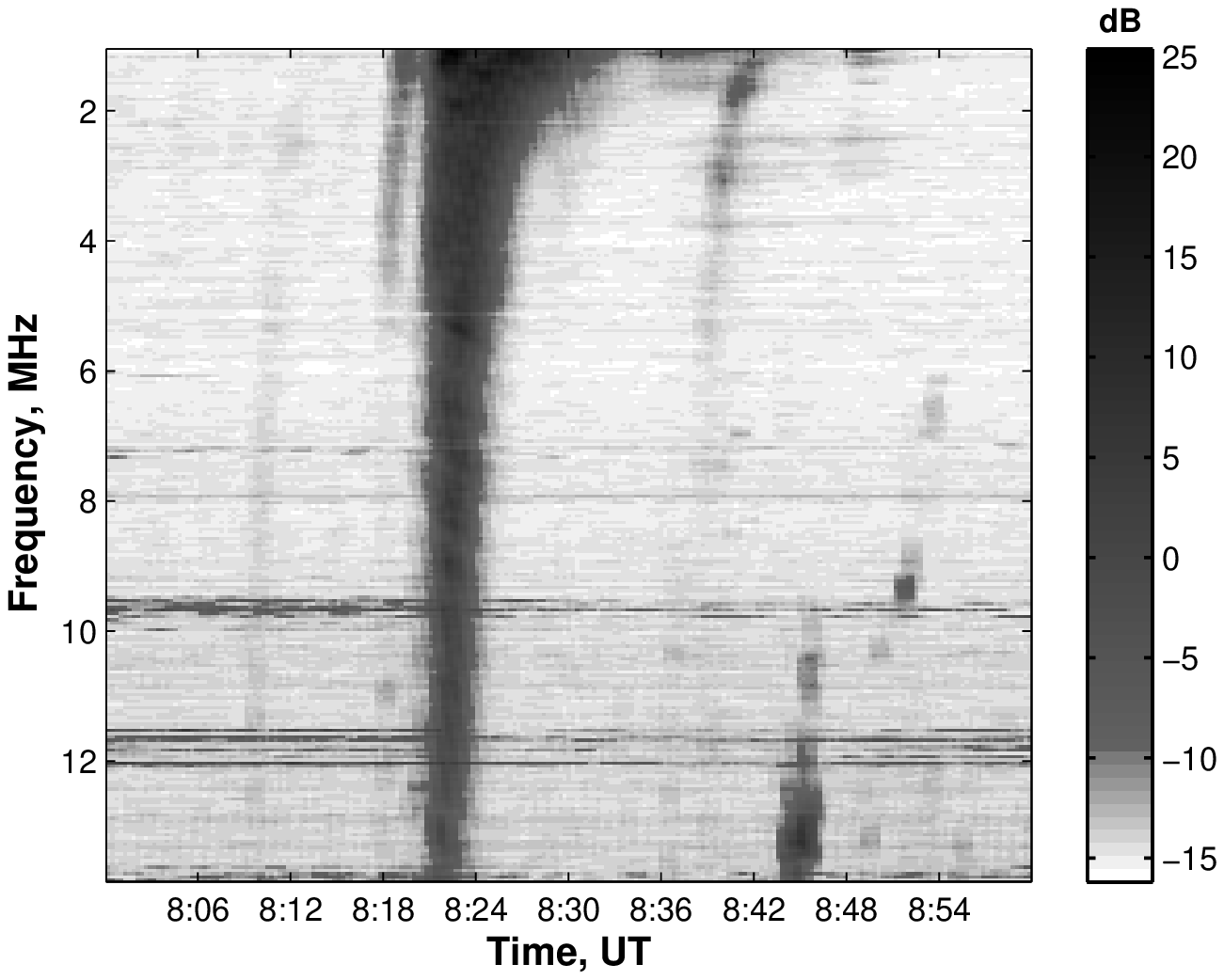}
\includegraphics[width=15cm]{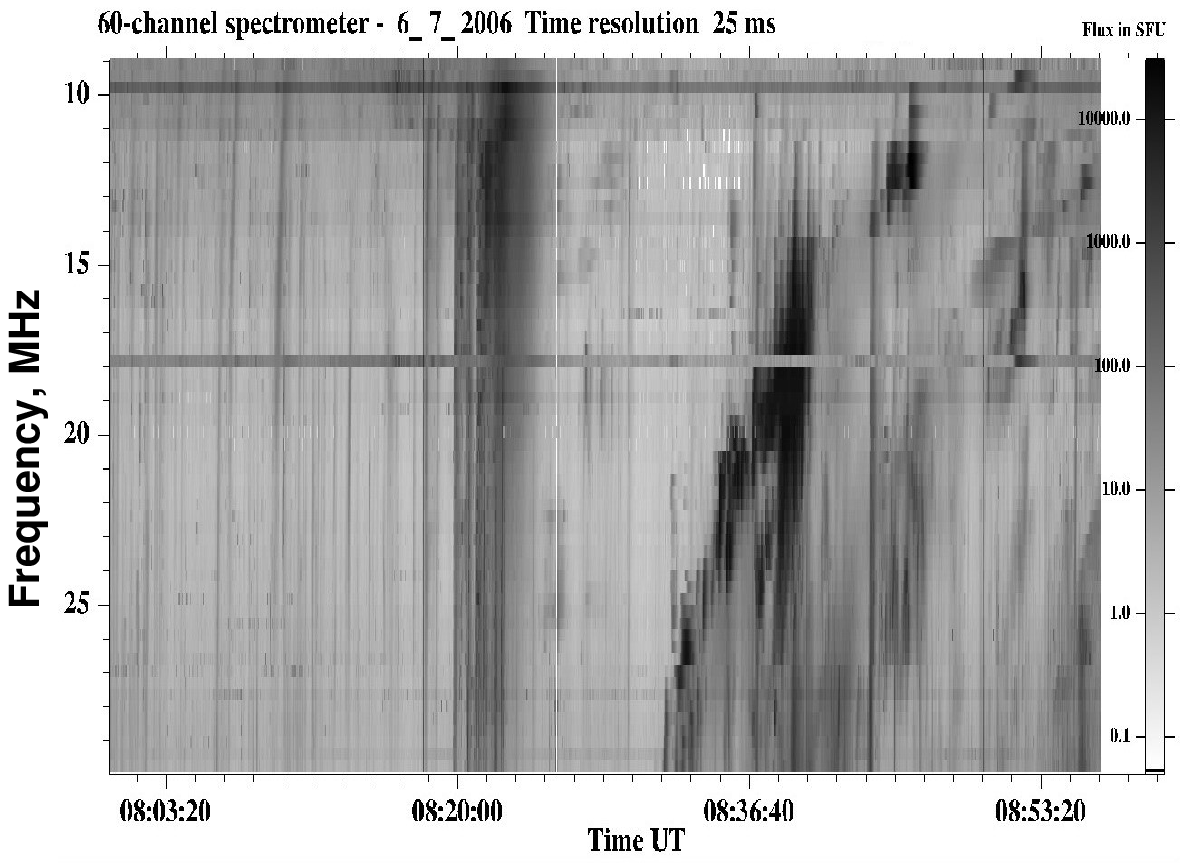}
 \caption{Set of solar bursts, observed in 2006 July 6 at 8:00-9:00 UT: bottom panel shows records
by the antenna array of 720 dipoles (UTR-2); top panel represents measurements of the
spacecraft WIND (both observations are simultaneous). 60-channel records of the UTR-2
array are expressed in terms of solar flux unity (sfu), and the WIND data are done with
respect to a background level.\label{fig3}}
\end{figure*}

We have chosen a number of solar events with different intensity features, that were
observed in the same time by the spacecraft WIND and the radiotelescope UTR-2, and
compared them between each other from the obtained records. The frequency band 12 MHz of
DSP, used now in the normal mode of UTR-2, is usually started with about 18.5 MHz because
of troubles with a sufficiently strong interference at lower frequencies. To save the
overlapping in the frequency band, we have decided to analyze the data of solar events,
measured by the 60-channel spectrograph (its frequency range 9-30 MHz is covered by 60
frequency channels). Its features are written in the paper of Mel'nik et~al. (2004)
together with DSP ones. In Fig. \ref{fig3} there is an illustration in point. It shows a
registration of solar bursts that were observed in 2006 July 6 simultaneously by both
instruments. The spectra distinctly indicate that the most intensive fragments of this
event can be only detectable for the Waves WIND instrument from its records. As for the
measurements, received by the radiotelescope UTR-2, they demonstrate a more detailed
structure in the event. So, from its scan it is seen up to ten different bursts. The
sensitivity and the temporal resolution of the receiver RAD2 for the fine analysis of
such events is insufficient. The fine structure is not distinguishable. In fact, from the
scan of the Waves WIND instrument at 9-13.8 MHz one can find out the type III burst at
8:20 UT and a fragment of the cloud structure of the type II burst at 8:45 UT. In the
same time from the scan, obtained by UTR-2, one can succeed in finding various details of
the given event, about which we could only assume from the WIND data. As another example,
it should be mentioned about interesting results reported in Konovalenko et al. (2007).
The matter is the absorption burst observed in 2003 August 19. This event is enough
difficult to detect from the corresponding registration of the Waves WIND instrument.
Without the data of the telescope UTR-2 it is doubtful whether its exact identification
was possible. Often we observe solar bursts having tracks on both UTR-2 and WIND records,
but there are also records on which UTR-2 detects clearly solar events without any
appreciable WIND verification in the overlapping range of frequency. Here it will be
useful to return to the antenna performance analysis again.

\section{Performance Comparison}
The antennas of the Waves WIND instrument are short dipoles in much of their respective
frequency range. Let us determine a performance of such an antenna for receiving the
decameter solar radio emission. In this case the system noise temperature $T_{sys}$ is
produced basically by the galactic background radiation. The galactic emission has been
studied for a long time by a many observers (see Krymkin 1971; Cane 1979; Manning and
Dulk 2001; Dulk et~al.  2001 and references therein). Cane (1979) suggested to describe
the flux density of the galactic background radiation in beam (measured in units of
W$\cdot$m$^{-2}\cdot$Hz$^{-1}\cdot$ sterad$^{-1}$) at the frequency range 1-100 MHz in a
clear functional form with a maximum at the frequency about 3 MHz. Her expression may be
approximated to within 2.4~\% at the frequency range 10-100 MHz by a simpler expression
\begin{equation}
I_\nu=I_{\rm gal}\,\nu^{-0.52}+I_{\rm ext}\,\nu^{-0.8}, \label{eq2}
\end{equation}
where $\nu$ is the frequency in MHz. The first term describes the galactic contribution,
and the second corresponds to the extragalactic radiation. In this equation the
parameters are equal to
\begin{equation}
I_{\rm gal}=2.48\cdot 10^{-20},\quad I_{\rm ext}=1.06\cdot 10^{-20}\,. \label{eq3}
\end{equation}
Of course, the flux density of the galactic background radiation also depends on the
solar location in the sky. Over large solid angles away from the galactic plane the flux
density $I_{\rm gal}$ is fairly constant and Eq.(\ref{eq2}) is applicable. However, near
the galactic plane an enhancement is superposed on the isotropic component and displays a
much richer structure in angle (see an interesting analysis of this problem in the paper
of Manning and Dulk (2001). The sky map may be divided into the two flux density regions
using galactic coordinates, as it was shown in Joardar and Bhattacharya (2007), but in
our study we will not account for the structure of galactic background radiation. We only
have restricted by Eqs. (\ref{eq2}) and(\ref{eq3}) in this consideration. Using the
Rayleigh-Jeans Law, it is not difficult to calculate the brightness temperature of the
galactic background radiation at 9-13.8 MHz. In particular, for the frequency 25 MHz it
is about 28300 K that comes to agreement with the measurements reported in the works of
Krymkin (1971) and Konovalenko et al.(2007).

The beam solid angle of the WIND-spacecraft antenna is $\Omega_{\rm ant}=8\pi/3$ sterad
(assuming, it receives in a homogeneous medium of infinite extent). Consistent with Kraus
(1967), its effective area equals $A_0=\lambda^2\times 3/8\pi$, where $\lambda$ is the
wavelength of radio emission. As a measure of the telescope performance, we apply the
System Equivalent Solar Flux Density (SESFD) similar to the ordinary definition of SEFD
(Campbell 2002). Consider an 1.0 sfu point source, and the antenna temperature due to
this source is expressed in terms of the K/sfu value defined in Eq. (\ref{eq1}). The
SESFD is the point source flux density (in sfu) which produces the antenna temperature
$T_{\rm A}= T_{sys}$, namely
\begin{equation}
{\rm SESFD}=T_{sys}/p. \label{eq5}
\end{equation}
For the antenna array of 720 dipoles with the sensitivity equal to 170,000 K/sfu and the
system noise temperature 130,000 K at 13.8 MHz, the SESFD is less 1 sfu, whereas for the
WIND-spacecraft antenna with the sensitivity 204 K/sfu at the same frequency, the SESFD
amounts to 637 sfu. Although the brightness temperature of galactic background radiation
is almost the same for receiving by either one dipole or the array, but the brightness
temperature of solar radio bursts will be directly proportional to the effective area of
the receiving antennas. In this comparative analysis we do not account for any losses in
the antennas themselves and their feeder systems, as well as the signal-to-noise ratio
per beam due to the back end. Nevertheless, this is quite enough to compare the
performance of these instruments. Strictly saying about the system noise temperature for
the antenna array of 720 dipoles, it should be also pointed out an appreciable (in
comparison with galactic background radiation) contribution of quiet solar emission. In
these conditions, due to the quiet solar emission, the system noise temperature increases
by about 30-40\% in the dependence of frequency.

It should be also mentioned about the difficulties of decameter radio observations for
any ground-based dipole in the daytime because of high levels of terrestrial interference
that are represented in the bottom panels of Figs. \ref{fig1} and \ref{fig2}. Due to the
selectivity of the UTR-2 array of 720 dipoles, it has proved to be a most valuable
instrument for the observing of solar radio bursts in such nonsimple conditions of
observation. There is also one point that it should be remembered. In the frequency range
9-14 MHz the solar events often happened with the duration on the order of tens seconds
and less (Abranin et~al. 1979; Mel'nik et~al. 2004; Mel'nik et~al. 2005a;b; Chernov
et~al. 2007; Konovalenko et~al. 2007), and therefore the temporal resolution of the
receiver RAD2 of the spacecraft WIND in 16 s is too rough for their analysis.

\section{Discussion and conclusions}
Sum up, it should be pointed out some advantages and disadvantages of space observations
of decameter radio emission from the solar corona in comparison with ground-based
measurements. The evident advantage is that the space observations permit ones to carry
out them continuous in time and beyond the transparency bounds of ionosphere (lower 7-9
MHz) that is problematic for ground-based instruments. Although in the former case one
can hope that if decameter radiotelescopes with a sufficient effective area were as many
as possible, and they were situated in different places of the Earth, then it would be
possible to organize a consistent patrol of solar events to be passed around the
observatories. At that time such a method of observations would become almost continuous
in time. As weak points of the spacecraft observations, it should be noticed problems to
launch a sufficient effective antenna for the decameter wavelength range, but any
second-rate antenna distinctly reduces to the benefits given by the space observatory due
to its space location. As a result, the sufficient strong events can be detected, but any
fine structure is washed out. But it is difficult to have a clear concept about any solar
event without accurate records. In this context the ground-based radiotelescope of type
UTR-2 has a clear advantage. Now a new large decameter radiotelescope, known as LOFAR
(LOw Frequency ARray), is being built in Holland. Its appearance is very important. The
point is that in the world there are a number of small antennas (consisting of one or
some dipoles) for the decameter radio astronomy purpose such as the antenna BIRS of
Erickson in Tasmania (Erickson 1997), antenna GBSRBS in NRAO
(http://www.astro.umd.edu/$\sim$white/gb/), the Culgoora Radiospectrograph (Prestage
et~al. 1994), the antenna of IZMIRAN at 25 MHz in Russia (http://helios.
izmiran.troitsk.ru/lars/LARS.html) and others. However, they have the same disadvantage
that we noticed above in respect to a too small effective area of such antennas. No
wonder that the instruments do not register weak solar events in the decameter wavelength
range, but this does not mean their lack. Sometimes, the impossibility of their detection
because of a small antenna aperture may be fatal for the analysis of solar events. Thus,
such kind of solar events is desirable to study by more comprehensive antennas with
effective back-ends. In closing it should be added that for the firm belief we applied
the aforesaid analysis for the data, represented in the internet site from the
measurements, obtained by the instrument GBSRBS. Though the instrument GBSRBS is located
in the North America, and UTR-2 is in East Europe, some solar events were observed
simultaneously by both radiotelecopes. It should be pointed out that their frequency
bands are overlapping. The comparison of the data has confirmed our above-mentioned
conclusions. All this does not means that we have something against the development of
one-dipole radioastronomy and space missions, and rather we want to note their real
performance capabilities. The importance of such one-dipole radio telescopes can scarcely
be overestimated for monitoring of solar bursts in Radio and Space Services. To each its
own. In this context we believe that the ground-based support by large radioastronomy
instruments will be very useful for the space missions in the sequel.

\section*{Acknowledgments}

The WAVES instrument is a joint effort of the Paris-Meudon
Observatory, the University of Minnesota, and the Goddard Space
Flight Center.

\label{lastpage}
\end{document}